\documentclass[pra,reprint,longbibliography
]{revtex4-1}
\usepackage{amsmath,amssymb,graphicx}
\usepackage{braket}
\usepackage{mathrsfs}
\begin{document}

\title{Complementary weak-value amplification with concatenated postselections}

\author{Gerardo I. Viza $^{1,3}$}
\author{Juli\'{a}n Mart\'{i}nez-Rinc\'{o}n$^{1,3}$}
\email{Corresponding author: jmarti41@ur.rochester.edu}
\author{Wei-Tao Liu$^{2,1,3}$}
\author{John C. Howell$^{1,3,4,5}$}

\address{$^1$Department of Physics and Astronomy, University of Rochester, Rochester, New York 14627, USA}
\address{$^2$College of Science, National University of Defense Technology, Changsha 410073, China}
\address{$^3$The Center for Coherence and Quantum Optics, University of Rochester, Rochester, New York 14627, USA}
\address{$^4$Institute of Optics, University of Rochester, Rochester, New York 14627, USA}
\address{$^5$Institute for Quantum Studies, Chapman University, Orange, California 92866, USA}

\begin{abstract}
We measure a transverse momentum kick in a Sagnac interferometer using weak-value amplification with two postselections. The first postselection is controlled by a polarization dependent phase mismatch between both paths of the interferometer and the second postselection is controlled by a polarizer at the exit port. By monitoring the darkport of the interferometer, we study the complementary amplification of the concatenated postselections, where the polarization extinction ratio is greater than the contrast of the spatial interference. In this case, we find an improvement in the amplification of the signal of interest by introducing a second postselection to the system.

\end{abstract}

\maketitle

\section{Introduction}\label{Intro}


Weak measurements were introduced by Aharonov, Albert and Vaidman~\cite{Aharonov}, and have proven a valuable tool for observation of quantum phenomena~\cite{Lundeen,Resch,Hardy,Lundeen2,Greg,QuantumCat,CheshireCatBrazil}. Weak-value amplification, a metrological technique for parameter estimation~\cite{Hosten,Dixon2009,Viza,Egan,StarlingFreq,jayaswal,Li,StarlingPhase}, has been shown to saturate the shot-noise limit~\cite{Viza} by imparting the precision of $N$ photons into a small subset of postselected photons~\cite{Jordan2013,Viza2}. The weak-value amplification technique has also been shown to exploit the geometric configurations of the experiment to reduce technical noise~\cite{Viza2,Jordan2013}. 

The weak-value amplification protocol starts with a well defined initial state of a system, $\ket{\varphi_i}$,  followed by an interaction given by $\hat{U}=\exp(-ig\hat{A}\hat{x})$ that couples the system, $\hat{A}$, to a continuous degree of freedom or a meter, given be $\hat{x}$. The interaction strength between the system and meter is given by parameter $g$, which we assume small. 
The system is then postselected to state $\ket{\varphi_f}$, which is near orthogonal to the preselected state $\ket{\varphi_i}$. The benefits for signal amplification and technical-noise mitigation require a weak system-meter interaction, and a strong postselection or data discarding on the meter. The mean value of the postselected outcomes in the meter is shifted by an amount proportional to the weak-value given by
\begin{equation}
A_{wv} = \frac{\bra{\varphi_f}\hat{A}\ket{\varphi_i}}{\braket{\varphi_f|\varphi_i}}.
\label{eq:wv}
\end{equation}

Weak-value amplification gives a metrological benefit based on an amplification of the signal relative to the technical-noise floor of the experiment~\cite{Viza2,Jordan2013}. Spatial interference experiments using a Sagnac interferometer as in Ref.~\cite{Dixon2009,Viza,Egan,StarlingFreq,StarlingPhase} have been limited to postselection angles of slightly under $5^\circ$, which is equivalent of discarding about 99$\%$ of the input photons in the interferometer. Others have studied regimes to maximize the amplification benefit of the technique by using a full theory without the linear approximation and finding a non-linear regime for very small postselection angles~\cite{Nakamura,PanOrthogonal,Koike,DiLorenzo}. 
Here we explore the complementary amplification by concatenating a second postselection to decrease the effective postselection angle to measure a beam deflection. 


In this work, we utilize the which-path information from a Sagnac interferometer and a polarization dependent phase offset between paths to measure a beam deflection. We concatenate two postselections: the first with spatial interference and the second with polarization interference. By monitoring the dark port of the sequence of postselections we record the beam shift proportional to the weak-value. We study the complementary amplification behavior between the spatial contrast and the polarization extinction contrast to the measured weak-value. We evaluate the technical difficulties of an imperfect interferometer by modeling the output of the interferometer for the single and concatenated postselection cases with a background parameter. We optimize the concatenated postselection and demonstrate a region of parameter space where the concatenated postselection provides an enhancement in the signal.

This paper is organized as follows. In Sec.~\ref{Sec:Theory} we start with the theory for single and concatenated postselection for weak-value amplification to measure beam deflection with a model that includes spatial interference imperfections. Then in Sec.~\ref{Sec:Experiment} we describe the experimental setup. In Sec.~\ref{Sec:Concatenated}, we present the result of the weak-value techniques. In Sec.~\ref{Sec:Info} we include a brief description of the theory of the relative Fisher information from the interferometer. Lastly, we discuss the results and conclude in Sec.~\ref{Sec:Conclusion}.

\begin{figure}[t]
 \centering
\includegraphics[scale=0.37]{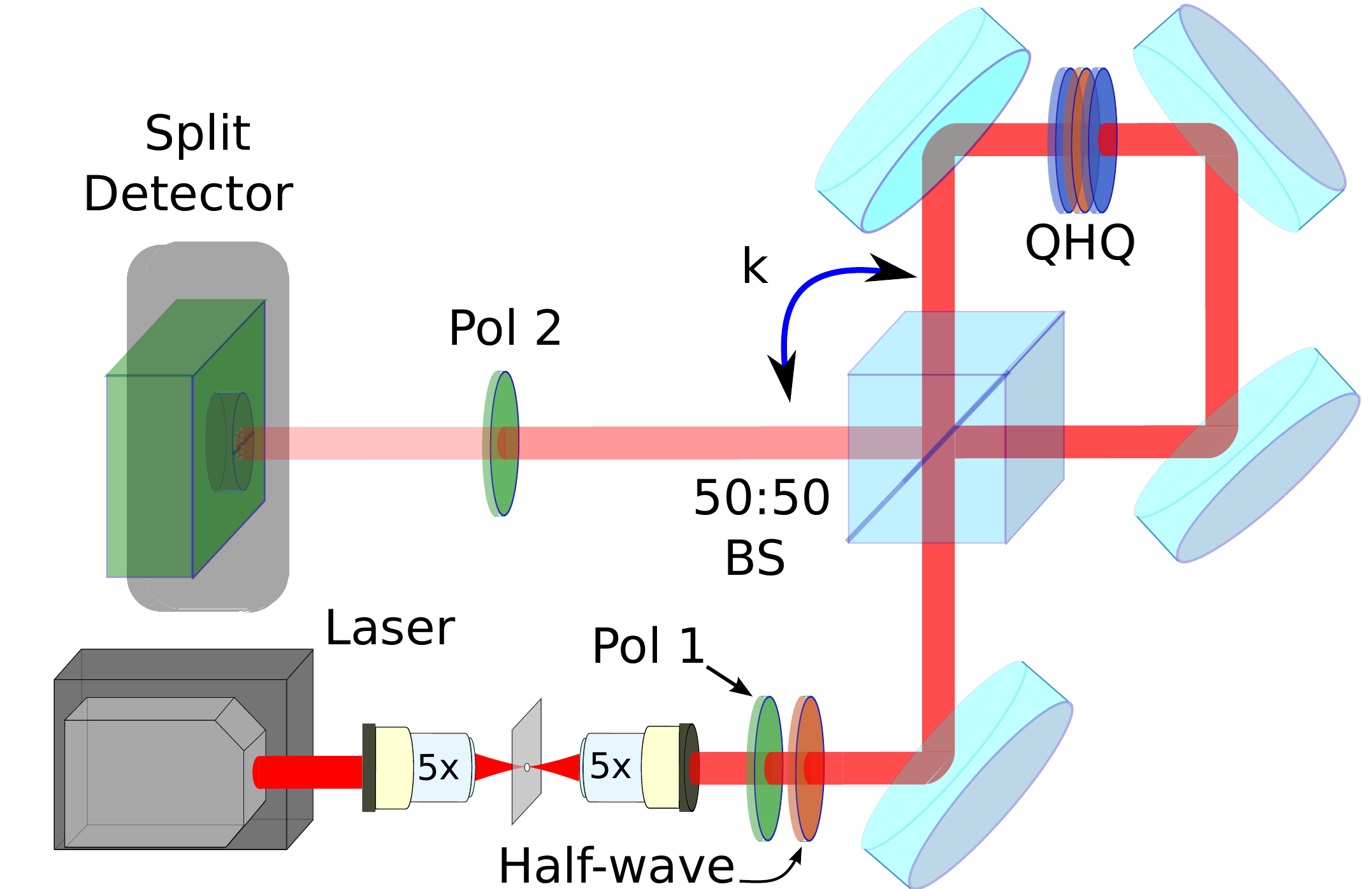}
 \caption{We send anti-diagonal polarized light (Pol $1$) through a Sagnac interferometer. Inside the interferometer are three wave plates arranged quarter-half-quarter (QHQ) designed to control the phase for clockwise and counterclockwise propagation (see App.~\ref{App:QHQ}).
The counterclockwise propagating beam receives a transverse momentum kick $k$ from the piezo-controlled $50$:$50$ beam splitter. When the beam recombines it destructively interferes at the dark port. The beam is then postselected a second time with polarizer $2$ (Pol 2). The half-wave plate after the first polarizer (Pol 1) is used for independent calibration of the horizontal and vertical polarization components.}
 \label{fig:SWVsetup}
\end{figure}

\section{Theory}\label{Sec:Theory}
A laser beam with a TEM$_{00}$ mode and $1/e^2$ beam radius $2\sigma$
enters a Sagnac interferometer through a piezo-actuated $50$:$50$ beam splitter (see Fig.~\ref{fig:SWVsetup}). The reflected beam receives a transverse momentum kick $k$ upon both entering and exiting the interferometer. We monitor the spatial beam shift of the beam exiting the dark port. The quarter-half-quarter (QHQ) wave plates combination~\cite{Hariharan} gives a Pancharatnam-Berry phase~\cite{Loredo} of $\pm\phi/2$ to each counter-propagating ($\ket{\circlearrowright},\ket{\circlearrowleft}$) beam (details are shown in App.~\ref{App:QHQ}). 
The interaction with the meter-system is given by  $\exp(-i k \hat{A} \hat{x})$, where the ancillary system  operator is $\hat{A} = \ket{\circlearrowright}\bra{\circlearrowright}-\ket{\circlearrowleft}\bra{\circlearrowleft}$.

For the remainder of the paper, we will describe the experiment in the classical matrix formalism~\cite{JohnQM}.
For input anti-diagonal polarized light the output electric field takes the form
\begin{equation}
{\bf E}^{out}(x;\beta,\alpha) = \frac{E_0}{2\sqrt{2}}\,e^{-x^2/4\sigma^2}
\begin{pmatrix}
\beta\,e^{i\alpha}-e^{-i2kx-i\phi}\\
\beta\,e^{i\alpha}-e^{-i2kx+i\phi}
\end{pmatrix},
\label{eq:EoutSingle}
\end{equation}
in the horizontal, $H=(1,0)^T$, and vertical, $V=(0,1)^T$, polarization basis. We introduce a spatial interference background using the relative amplitude $\beta$ and the relative phase $\alpha$. The constant $\beta$ describes the relative transmission amplitude through both arms in the interferometer, which limits the spatial contrast from reaching the perfect zero output from the dark port and the perfect input power in the bright port. The introduction of the phase $\alpha$ allows us to make a correct estimate of $\phi$, which accounts for the effective amplification in the parameter $k$. The free parameters $\beta$ and $\alpha$ in the theory account for background light due to imperfect alignment, imperfections of the optical elements in the experimental setup, and systematic errors in the estimates of $k$. These parameters are expected to take values $\beta=1$ and $\alpha=0$ in a perfect interferometer.
The difference in sign for $\phi$ in Eq.~(\ref{eq:EoutSingle}) for $H$ and $V$ polarization comes from the asymmetric response of the QHQ combination inside the interferometer (see App.~\ref{App:QHQ}).

\subsection{Theory: Single Postselection}
First we assume the ideal case of $\beta=1$ and $\alpha=0$. The polarization degree of freedom will be used for the second postselection, so we assume here that the input light is either horizontally or vertically polarized.  Using the modulus square of one component of the electric field from Eq.~(\ref{eq:EoutSingle}) we arrive with the intensity profile. With the intensity profile we assume the momentum kick is small for the
weak interaction approximation, $k^2\sigma^2\cot^2(\phi/2)\ll1$. We expand the trigonometric functions up to first order in $k$ and re-exponentiate the quantity. Then we combine the two exponentials by completing the square to arrive at the dark port intensity profile,

\begin{equation}
I_s(x)=I_0\sin^2\left(\frac{\phi}{2}\right)\exp\left[-\frac{1}{2\sigma^2}(x-\delta x_s)^2\right].
\label{eq:dxsingle}
\end{equation}
The subscript $s$ of Eq.~(\ref{eq:dxsingle}) refers to the single postselection where $\delta x_s=\pm2k\sigma^2\cot(\phi/2)$ is the beam shift from a horizontally or vertically polarized input light. This is the standard result from the beam deflection experiment~\cite{Dixon2009}, with a weak-value of $A_{wv} = \pm i\cot(\phi/2)$ (see App.~\ref{App:WVQM}).

For realistic experimental implementations when $\phi$ is small we assume the case of $\beta<1$ and $\alpha\neq0$ in Eq.~(\ref{eq:EoutSingle}). Integrating the square modulus of the electric field of either $\ket{H}$ or $\ket{V}$ of Eq.~(\ref{eq:EoutSingle}) yields the normalization factor
\begin{equation}\label{Eq:Ns}
\mathcal{N}_s=\beta\,\sin^2\left(\frac{\alpha\mp\phi}{2}\right)+\left(\frac{1-\beta}{2}\right)^2.
\end{equation}
The mean beam shift on the detector is then given by
\begin{equation}
\begin{split}
\braket{x}_s &= \frac{1}{\mathcal{N}_s}\int x |E_{H,V}^{out}(x;\beta,\alpha)|^2 dx\\
&=\frac{\beta k \sigma^2\,\sin(\alpha\mp\phi)}{\mathcal{N}_s}.
\end{split}
\label{eq:swvCorrection}
\end{equation}
Note that the mean shift in Eq.~(\ref{eq:swvCorrection}) has two solutions that depend on the different components of Eq.~(\ref{eq:EoutSingle}).

\subsection{Theory: Concatenated Postselection}
The second part of the theory is to take advantage of the polarization sensitive phase $\phi/2$ by inputting 
anti-diagonal polarized light as in Eq.~(\ref{eq:EoutSingle}). 
For the ideal case of $\beta=1$ and $\alpha=0$, the orthogonal components of polarization will spatially separate at the dark port by $2|\delta x_s|$ since the horizontal and vertical components have opposite weak-values (see App.~\ref{App:WVQM}).
The electric field exits the Sagnac interferometer and passes through a polarizer with a Jones matrix given by
\begin{equation}
\bold{P}(\theta)=\frac{1}{2}
\left({\begin{array} {c c}
1+\sin(2\theta)& \cos(2\theta)\\
\cos(2\theta) & 1-\sin(2\theta)
\end{array}}
\right).
\label{Eq:pol}
\end{equation}
The polarizer angle $\theta$ is aligned to be nearly orthogonal to the polarization of the exit beam from the interferometer.

We assume the momentum kick is small for the weak interaction approximation, $k^2\sigma^2\cot^2(\phi/2)\cot^2(\theta)\ll1$. We expand the trigonometric functions up to first order in $k$ and re-exponentiate the result. We then combine the exponentials by completing the square
to arrive at the dark port intensity profile,
\begin{equation}
I_{c}(x)=I_0\sin^2\left(\frac{\phi}{2}\right)\sin^2(\theta)\exp\left[-\frac{(x-\delta x_c)^2}{2\sigma^2}\right].
\label{eq:IntensityConc}
\end{equation}
The beam shift after the concatenated postselection is given by $\delta x_{c}=2k\sigma^2\cot(\phi/2)\cot(\theta)=\delta x_s\cot\theta$. The subscript $c$ refers to the concatenated postselected case.

We now assume the realistic case where $\beta<1$ and $\alpha\neq0$.
After the polarizer we have a new normalization term
\begin{eqnarray}\label{eq:Nc}
 \mathcal{N}_c&=&\frac{1+\beta^2}{8}\left[1+\cos(2\theta)\cos\phi\right]\nonumber\\
 &+&\frac{\beta}{4}\left[\sin\alpha\sin\phi\sin(2\theta)-\cos\alpha(\cos(2\theta)+\cos\phi)\right],
\end{eqnarray}
and a new mean shift of the beam on the detector
\begin{equation}
\begin{split}
\braket{x}_c  &= \frac{1}{\mathcal{N}_c}\int x|{\bf P}(\theta){\bf E}^{out}(x;\beta,\alpha)|^2 dx\\
&=\frac{\beta k\sigma^2}{2\mathcal{N}_c}\left[\sin\alpha(\cos(2\theta)+\cos\phi)+\cos\alpha\sin(2\theta)\sin\phi\right].
\end{split}
\label{eq:dwvCorrection}
\end{equation}
Note that by setting $\beta=1$ and $\alpha=0$ we recover the expressions $\mathcal{N}_c\rightarrow\sin^2(\phi/2)\sin^2(\theta)$ and $\langle x\rangle_c\rightarrow \delta x_c$ in Equations~(\ref{eq:Nc}) and~(\ref{eq:dwvCorrection}). 


\section{Experiment}\label{Sec:Experiment}
The experimental setup shown in Fig.~\ref{fig:SWVsetup} starts with a grating feedback laser with a $780$ nm center wavelength. Two objectives and a $50$ $\mu$m pinhole are used to create a collimated Gaussian beam with radius $2\sigma$, and a polarizer (Pol 1) selects the anti-diagonal linear polarization for the experiment.
The beam enters the Sagnac interferometer through a $50$:$50$ beam splitter on a piezo-actuated mount that provides a transverse momentum kick $k$ at each reflection. The interferometer has three wave plates, QHQ, that give a phase difference between paths. The quarter-wave-plates are set to $+45^\circ$ and $-45^\circ$. The half-wave plate sets the added phase of $\pm\phi/2$ to each path (see App.~\ref{App:QHQ}). When the beams recombine, they destructively interfere at the dark port. We monitor the beam shift of the light exiting the dark port with a split detector. In the second part of the experiment, we add a polarizer before the detector for the concatenated postselection (Pol 2 in Fig.~\ref{fig:SWVsetup}).

We use a beam radius of $2\sigma=1100$ $\mu$m with a polarization extinction ratio of $25000$:$1$. 
The polarization quality of the interferometer is limited to $5000$:$1$ by the wave plate combination inside the interferometer (QHQ in Fig.~\ref{fig:SWVsetup}). The traverse momentum kick $k$ is driven by a piezo stack calibrated separately for $100$ Hz with a response of $\alpha\approx63.9$ nm/V. For all the measurements, we apply a $100$ mV sinusoidal wave to the piezo stack which corresponds to a momentum kick $k=2.74$ m$^{-1}$. The first postselection angle $\phi/2$ is determined by the ratio of the measured power at the dark port, $P_{\phi/2}$, to the power of the bright port, $P_{bright,1}$, given by $P_{\phi/2}=\sin^2(\phi/2)P_{bright,1}$. The second interference postselection angle $\theta$ is determined by the ratio of the power after the output polarizer dark port, $P_{\theta}$, (Pol $2$ in Fig.~\ref{fig:SWVsetup}) to the power of the polarization interference bright port, $P_{bright,2}$, as in $P_{\theta}=\sin^2(\theta)P_{bright,2}$. 

\section{Results}\label{Sec:Concatenated}
%

\subsection{Single Postselection}
\begin{figure}[t!]
 \centering
 \includegraphics[scale=0.11]{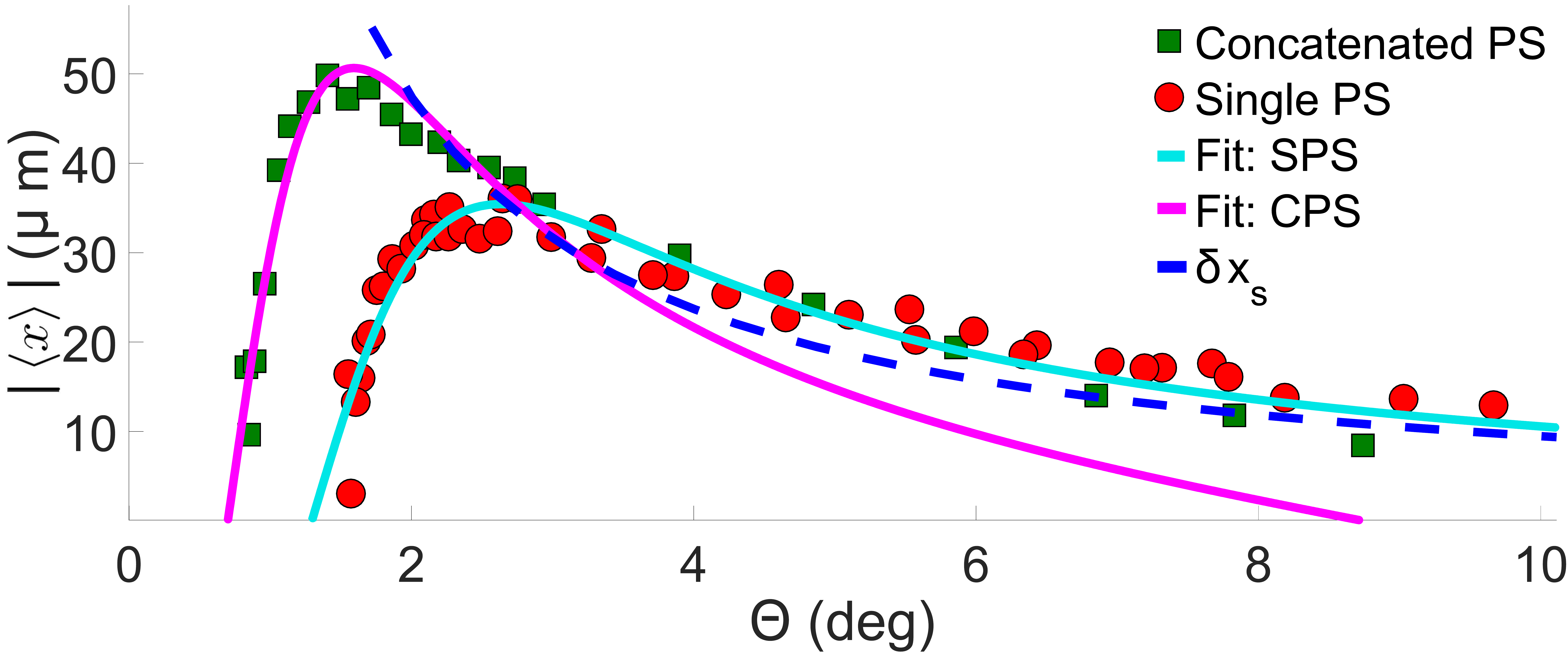}
 \caption{The average beam shift as a function of postselection angle $\Theta$. The variable $\Theta$ is a generalized postselection angle; for the single postselection case, $\Theta=\phi/2$ (red circles), but for the concatenated postselection case it is the product of both postselection angles $\Theta=\theta\phi/2$ (green squares). The label PS refers to postselection. The theory (dotted blue line) is labeled as $\delta x_s=4k\sigma^2/\Theta$ as in the mean beam shift of Eq.~(\ref{eq:dxsingle}) with generalized postselection angle $\Theta$. Fit: SPS (solid teal line) is the fit of the single postselection data, and Fit: CPS (solid purple line) is the fit to the concatenated postselection data as in Eqs.~(\ref{eq:swvCorrection}) and~(\ref{eq:dwvCorrection}), respectively. 
We present here the concatenated data set from row three in Table~\ref{tab:table1}.}
 \label{fig:SWVbeta}
\end{figure}

In Fig.~\ref{fig:SWVbeta}, we plot (red circles for single postselection) the absolute mean value of the beam shift versus the generalized postselection angle $\Theta$. The generalized postselection angle for the single postselection case is given by $\Theta=\phi/2$. 
The data consists of both horizontal and vertical polarized input light with dark port contrast of $1400$:$1$. From Eq.~(\ref{Eq:Ns}), the dark port contrast ratio is given by $(1-\beta)^2/(1+\beta)^2$, so a value of $\beta\approx0.95$ is expected. However, we numerically fit the data to Eq.~(\ref{eq:swvCorrection}) and label $\beta$ and $\alpha$ as free parameters. The fit to the single postselection data is labeled as Fit: SPS (solid teal line) and takes on the positive values of $\delta x_s$ as in the horizontal polarized case of Eq.~(\ref{eq:swvCorrection}). From Fit: SPS we extract the optimal 
postselection angle $\Theta_{opt}$, where the weak-value amplification shows the largest shift before the signal is overcome by the background for small postselection angles. From Fit: SPS we observe that the largest signal is found with a postselection angle of $\phi/2=2.6^\circ$.  We also see that the relative transmission amplitude parameter is given by $\beta=0.9543(8)$. The data differs slightly from the theory of Eq.~(\ref{eq:dxsingle}) (dotted blue line) because of a systematic error in calibration of the piezo-actuated beam splitter. This theory of Eq.~(\ref{eq:dxsingle}) is the beam shift $\delta x_s = 2k\sigma^2\cot(\phi/2)$ without the spatial interference imperfection consideration. 


We note that as the alignment improved, the quality of the dark port also improved and the optimal angle for greatest amplification decreased. The results from the single postselection case of Fig.~(\ref{fig:SWVbeta}) show that the relative amplitude transmission $\beta<1$ limits the weak-value amplification from the theoretical upper bound~\cite{PanOrthogonal,ShengOrthogonal} and limits the benefit over technical noise~\cite{Jordan2013,Viza2}.

\subsection{Complementary Behavior Between Postselections}
We note the complementary behavior of the two degrees of freedom used for postselection, which-path and polarization. For example, if one postselects the spatial interference to resolve maximum amplification of the single postselection (the peak of Single PS in Fig.~\ref{fig:SWVbeta}), then there cannot be any polarization improvement because we observe a extinction contrast close to $10$:$1$ for polarization. To understand this limitation we first note that
if the first postselection output power is $P_{\phi/2}$, the contrast ratio is $P_{\phi/2}$:$P_{in}$ and our case of maximum amplification in the Single PS gives $P_{\phi/2}$:$P_{in}\approx500$:$1$. Then we observe the second postselection to have a contrast at best of about $10$:$1$ for a total effective contrast of $5000$:$1$. The effective contrast of both postselections cannot exceed the contrast of either the spatial or polarization extinction contrast. 
In this particular scenario, we are not in the small angle (because of $10$:$1$ in polarization extinction contrast) regime so the configuration is suboptimal. We note with a maximum polarization extinction contrast of $10$:$1$ we can expect the location of the peak slightly close to $\theta\approx 0.5$ rad $\approx 29^\circ$. This angle will not provide an enhanced beam shift of $\delta x_c$, as the angle is outside the small angle approximation and will not follow the optimal theory of Eq.~(\ref{eq:IntensityConc}).
For our experiment, the spatial interference contrast is $1400$:$1$ and the polarization contrast is $5000$:$1$. 
Therefore, we present the following results of the \emph{optimized} case in the next section.

\subsection{Concatenated Postselection}
When the single postselected beam's shift is compared to the concatenated case, we see from Eq.~(\ref{eq:IntensityConc}) that the beam shift is amplified by $\cot(\theta)$ at the cost of a fraction $\cos^2(\theta)$ of less measurements. 

We point out that the theory of Eq.~(\ref{eq:IntensityConc}) does not assume any limitations to the contrast for either spatial or polarization interference. In the case of infinite contrast there is no benefit in adding a second postselection. Since this is an idealization, therefore we explore the case of having one degree of freedom with a higher contrast than the other. In this experiment, the spatial interference contrast is $1400$:$1$ and polarization contrast is $5000$:$1$, thus there exists an optimal configuration for the complementary amplification.

Now we focus on the concatenated postselection data (green squares) in Fig.~\ref{fig:SWVbeta}. 
We plot the absolute value of the mean beam shift, $|\braket{x}|$, from Eq.~(\ref{eq:IntensityConc}) as a function of the generalized postselection angle $\Theta$, which takes the form $\Theta=\theta\phi/2$ for the concatenated postselection. The product of postselection angles is a valid
approximation for the small angle regime. The plot shows the benefit of 
introducing the polarization degree of freedom to the experiment which allows us to 
achieve smaller effective postselection angles $\Theta$ and larger shifts $\langle x\rangle$. From Fit: CPS (solid purple line) from Eq.~(\ref{eq:dwvCorrection}), the optimal postselection 
angle for the concatenated postselection is about $1.6^\circ$. 
From the fits shown in Fig.~\ref{fig:SWVbeta} the improved beam shift with the concatenated postselection has increased by a factor of approximately $50.64/35.425\sim1.4$ over the single postselection beam shift. The fit of the concatenated weak-value case gives a background interference parameter 
$\beta=0.968(1)$~\footnote{We remind the reader that the data presented for the concatenated case is one of the two optimized cases (rows 3 and 4)  where there is an enhancement in the signal, i.e. row 3 in Table~\ref{tab:table1}.}.

Now we compare the spatial interference background parameter $\beta$ of the single and concatenated cases. The fitting of Eq.~(\ref{eq:swvCorrection}) and Eq.~(\ref{eq:dwvCorrection}) to the data reveals $\beta\approx0.9543(8)$ and $\beta\approx0.968(1)$, respectively. The error is from the fit of the data with $95\%$ confidence. By introducing the second postselection the parameter $\beta$ is increased, showing the advantage of the concatenation for interference improvement. We note that such an increase is present even after the introduction of  a non-ideal optical element (Pol 2 in Fig.~\ref{fig:SWVsetup}) which could in principle reduce the spatial interference.


\section{Fisher Information}\label{Sec:Info}
In this section, we theoretically compare the efficiency of the concatenated weak-value technique 
in the ideal noiseless case. 
We use the Fisher information formalism of the parameter of interest $k$ given by
\begin{equation}
\mathcal{I}(k)=\int dx P(x;k)\left[\frac{\partial}{\partial k}\ln{P(x;k)}\right]^2,
\label{Eq:FisherInfo}
\end{equation}
where $P(x;k)$ is the probability distribution of the photons arriving on the detector. 

\begin{table}
\caption{\label{tab:table1}Results of concatenated postselection for weak-value amplification. The results of the last three columns come from numerically fitting the data where the nonlinear fit is  with a $95\%$ confidence and a goodness measure $r^2>0.86$. The first column is the contrast of the first postselection. The second column is the first postselection angle, $\phi/2$. $\Theta_{opt.}$ from the third column is the product $\theta\phi/2$ with highest signal from the fit. The fourth column is the relative transmission amplitude parameter $\beta$ from the fit. The quantity $\mathcal{I}_c(k)_{frac}$ of column five is the fraction of Fisher information with respect to the total Fisher information considering both bright and dark ports of the first postselection. For column five, we consider Eq.~(\ref{eq:concatenatedInfo}) for rows two through five. The first row in column five has $\mathcal{I}_s(k)_{frac}\approx 1$ that corresponds to the fraction Fisher information of one postselection as in Eq.~(\ref{eq:singleInfo}). The proximity to one means that there is no loss of Fisher information when monitoring the dark port. This is not to be confused with shot-noise limited measurements because this is fractional Fisher information.}
\begin{tabular}{p{2.2cm}p{0.9cm}p{0.9cm}p{1.7cm}p{1.3cm}}

 $P_{bright,1}$:$P_{\phi/2}$& $\phi/2$ & $\Theta_{opt.}$ & \hspace{0.5cm}$\beta$ & $\mathcal{I}_{s,c}(k)_{frac}$\\
 \hline\hline
  \hspace{0.6cm}-- & \hspace{0.1cm}-- & $2.66^\circ$ & $0.9543(8)$& $\approx1.0$ \\
  \hline
 \hspace{0.5cm}$177:1$ & $4.33^\circ$ & $2.0^\circ$  & $0.962(1)$& $0.79$\\
 \hspace{0.5cm}$106:1$ & $5.56^\circ$ & $1.6^\circ$ & $0.968(1)$& $0.91$\\ 
 \hspace{0.5cm}$53.2:1$ & $7.88^\circ$ & $1.6^\circ$ & $0.965(1)$& $0.94$\\
 \hspace{0.5cm}$26.6:1$ & $11.2^\circ$ & $2.0^\circ$ & $0.964(1)$& $0.93$\\
 \hline
\end{tabular}
\end{table}

%

We consider the probability function of the optimized concatenated weak-value technique
\begin{equation}
P_{c}(x;k) = \prod^{N_c}_{i=1}\frac{1}{\sqrt{2\pi\sigma^2}}\exp\left[-\frac{(x_i-\delta x_c)^2}{2\sigma^2}\right],
\end{equation}
where $\delta x_c = 2k\sigma^2\cot^2(\phi/2)\cot^2\theta$ as in Eq.~(\ref{eq:IntensityConc}) is the complementary amplification of the concatenated postselection.
The probability function of the concatenated case has $N_c$ independent measurements which is less than the total number of possible measurements that are thrown away by the bright port. 
We also study the amount of Fisher information that is collected out of the dark port of each technique. We note the total available Fisher information is the sum of the Fisher information from the dark port (D) and the bright port (B) of the single postselection case, $\mathcal{I}_{s,D}+\mathcal{I}_{s,B}=4N\sigma^2$.
The Fisher information from the dark port of the single postselection and the concatenated postselection is written as
\begin{subequations}
\begin{equation}
\mathcal{I}_{s,D}(k) = 4N\sigma^2\cos^2(\phi/2),
\label{eq:Is}
\end{equation}
and
\begin{equation}
\mathcal{I}_{c,D}(k) = 4N\sigma^2\cos^2(\phi/2)\cos^2(\theta),
\label{eq:Ic}
\end{equation}\label{Eq:TotalFisherInfo}
\end{subequations}
respectively. The subscripts $s$ and $c$ refer to the single and concatenated cases respectively. 
The total number of possible measurements is $N$, however the single and concatenated techniques are limited to $N\sin^2(\phi/2)$ and $N\sin^2(\phi/2)\sin^2\theta$ measurements respectively.
The fractional Fisher information for the single postselection and concatenated case is given by 
\begin{subequations}
\begin{equation}
\mathcal{I}_s(k)_{frac} = \frac{\mathcal{I}_{s,D}}{\mathcal{I}_{s,D}+\mathcal{I}_{s,B}}=\cos^2(\phi/2),
\label{eq:singleInfo}
\end{equation}
and
\begin{equation}
\mathcal{I}_c(k)_{frac} = \frac{\mathcal{I}_{c,D}}{\mathcal{I}_{s,D}+\mathcal{I}_{s,B}}=\cos^2(\phi/2)\cos^2(\theta),
\label{eq:concatenatedInfo}
\end{equation}
\label{eq:FractionalFisherInfo}
\end{subequations}
\label{eq:Fractional}
\noindent
respectively.

\subsection{Results of the Comparison}
In Table~\ref{tab:table1}, we present the data from the single and the concatenated postselections. The first column is the output power of the spatial interference contrast. The first row is the single postselected case where all postselected values are measured. The single postselected case has no first or second column because it only uses spatial interference, where $\Theta=\phi/2$. The concatenated results are in the bottom four rows. The second column is the first postselection angle, $\phi/2$. The third column is the generalized postselection angle, $\Theta_{opt.}$, of the largest beam shift, $\delta x_c$, from the fits. The fourth column is the spatial interference background parameter $\beta$ from the fits. The fifth column is the fractional  Fisher information, $\mathcal{I}(k)_{frac}$ with respect to the total Fisher information of the system. For the first row of column five we consider the Fisher information as in Eq.~(\ref{eq:singleInfo}) and for the bottom four rows of column five we consider Eq.~(\ref{eq:concatenatedInfo}).

From Table~\ref{tab:table1}, the complementary amplification of the concatenated postselection is presented. We plot in Fig.~\ref{fig:SWVbeta} the optimized experimental run, given in row three of Table~\ref{tab:table1}. This optimized case shows not only the greatest amplification for the smallest postselection angle but also the lowest amount of spatial interference background, given by large $\beta$ in column four. 

The first row of Table~\ref{tab:table1} has fractional Fisher information given by Eq.~(\ref{eq:singleInfo}).
The proximity to $1$ of the fractional Fisher information in the first row means there is no loss in Fisher information due to only monitoring the dark port. This is not to be confused with a shot-noise limited measurement because this is a fractional description of the Fisher information meant to describe efficiency of the postselected events.


Looking at the third and fourth rows of Table~\ref{tab:table1}, the largest beam shift is found with a postselection angle of $\Theta_{opt}=1.6^\circ$. The single postselected angle $\Theta_{opt}=2.66^\circ$ has therefore improved with the optimized concatenated case (rows 3 and 4). We note that there exists an optimized case because the polarization degree of freedom has a greater extinction efficiency than the spatial degree of freedom in this experiment. 
The concatenated postselection could not be optimized any further because of limitations on our  polarization extinction ratio of $5000$:$1$.

An example of a non-optimized case for the second postselection is row 2 in Table~\ref{tab:table1}, where the angle for the first postselection $\phi/2$ is small ($4.33^\circ$). Such a selection for $\phi$ forces a large selection of $\theta\sim25.8^\circ$ which does not allow for greater peak amplification and decreases the idealized Fisher information by $21\%$. 
The loss of the available Fisher information is less than $9\%$ ($5\%$ loss in sensitivity) for the optimized region.
The optimized case can only exist when one interference contrast is higher than the other. Polarization is an example of a large extinction efficiency where polarizers can have extinction ratios of $10^6$:$1$. The introduction of a \textit{non-ideal} optical element (second polarizer in Fig.~\ref{fig:SWVsetup}) reduces the maximum amount of photons reaching the detector. For a maximum transmission of (typical) $85\%$ in the polarizer the shot-noise is increased by $8.5\%$ with respect to the single postselection scenario. This disadvantage is overshadowed by the amplification of $40\%$ obtained in the shift due to the second postselection in our technical-limited setup.


In this experiment, we postselect with spatial interference to produce two opposite weak-values (see App.~\ref{App:WVQM}) each of which carries half of the available Fisher information. Then we use the higher extinction contrast degree of freedom of polarization to postselect a second time to explore the complementary amplification between the two postselections. We find an optimized region of parameter space such that the complementary amplification is realized, and provides some benefit with a smaller effective postselection angle $\Theta$ and a decrease of the spatial interference imperfection (increase of parameter $\beta$).

It is worth pointing out the optics used in our experiment limited the polarization extinction
contrast to $5000$:$1$. This is consistent with our 
measurements of the concatenated postselection in Table~\ref{tab:table1}. 
We note that with higher performing optics we will amplify the signal 
 and circumvent spatial interference background. 
We also note this work is not to be confused with Ref.~\cite{Brun}, where they propose an entangled ancillary system to improve the precision of a measurement. In our experiment, the best precision possible is bounded by the shot-noise limit.

\section{Conclusion}\label{Sec:Conclusion}
In this paper, we have explored a complementary amplification of the concatenated post-selection for weak-values amplification to measure a beam deflection. We used a Sagnac interferometer with spatial interference to measure a transverse beam deflection and then introduced a second postselection to the system with polarization. The concatenated postselection angle, $\theta$, and the first spatial interference postselection angle, $\phi/2$, are complementary bounded by the highest interference contrast. Only when one of the two interference contrasts is larger than the other can there be an optimized regime in parameter space to observer the complementary amplification. 

In general, it is better to do one postselection, but in the case of low contrast spatial interference we can incorporate a higher contrast degree of freedom such as polarization for improvement. Thus from the optimized case the complementary amplification of concatenating postselections can lead to a more idealized interferometer according to a greater value of beta.
With higher quality optics we could have greater discrepancy between spatial and polarization extinction ratios and further increase postselection contrast in an optimized case. This condition would lead to greater reduction in technical noise~\cite{Viza2,Jordan2013} which would help reach the shot-noise limit with greater ease. It is worth noting that a concatenated postselection for weak-values is beneficial only when the additional degree of freedom has a higher interference contrast than the first interference. 

A new weak-values technique without postselection has recently been developed where the undesirable decay of signal for small angles of Fig.~\ref{fig:SWVbeta} is not observed~\cite{Julian,Weitao,Zekai}. This new technique produces an amplification to the signal of interest without the cost of reduced photon counts.

\begin{acknowledgments}
We give thanks for the careful editing by Bethany J. Little, Justin M. Winkler and Samuel H. Knarr. This work was supported by the Army Research Office Grant No. W911NF-12-1-0263, by the National Natural Science Foundation of China Grant No. 11374368 and the China Scholarship Council.
\end{acknowledgments}


\begin{thebibliography}{31}%
\makeatletter
\providecommand \@ifxundefined [1]{%
 \@ifx{#1\undefined}
}%
\providecommand \@ifnum [1]{%
 \ifnum #1\expandafter \@firstoftwo
 \else \expandafter \@secondoftwo
 \fi
}%
\providecommand \@ifx [1]{%
 \ifx #1\expandafter \@firstoftwo
 \else \expandafter \@secondoftwo
 \fi
}%
\providecommand \natexlab [1]{#1}%
\providecommand \enquote  [1]{``#1''}%
\providecommand \bibnamefont  [1]{#1}%
\providecommand \bibfnamefont [1]{#1}%
\providecommand \citenamefont [1]{#1}%
\providecommand \href@noop [0]{\@secondoftwo}%
\providecommand \href [0]{\begingroup \@sanitize@url \@href}%
\providecommand \@href[1]{\@@startlink{#1}\@@href}%
\providecommand \@@href[1]{\endgroup#1\@@endlink}%
\providecommand \@sanitize@url [0]{\catcode `\\12\catcode `\$12\catcode
  `\&12\catcode `\#12\catcode `\^12\catcode `\_12\catcode `\%12\relax}%
\providecommand \@@startlink[1]{}%
\providecommand \@@endlink[0]{}%
\providecommand \url  [0]{\begingroup\@sanitize@url \@url }%
\providecommand \@url [1]{\endgroup\@href {#1}{\urlprefix }}%
\providecommand \urlprefix  [0]{URL }%
\providecommand \Eprint [0]{\href }%
\providecommand \doibase [0]{http://dx.doi.org/}%
\providecommand \selectlanguage [0]{\@gobble}%
\providecommand \bibinfo  [0]{\@secondoftwo}%
\providecommand \bibfield  [0]{\@secondoftwo}%
\providecommand \translation [1]{[#1]}%
\providecommand \BibitemOpen [0]{}%
\providecommand \bibitemStop [0]{}%
\providecommand \bibitemNoStop [0]{.\EOS\space}%
\providecommand \EOS [0]{\spacefactor3000\relax}%
\providecommand \BibitemShut  [1]{\csname bibitem#1\endcsname}%
\let\auto@bib@innerbib\@empty
\bibitem [{\citenamefont {Aharonov}\ \emph {et~al.}(1988)\citenamefont
  {Aharonov}, \citenamefont {Albert},\ and\ \citenamefont
  {Vaidman}}]{Aharonov}%
  \BibitemOpen
  \bibfield  {author} {\bibinfo {author} {\bibfnamefont {Yakir}\ \bibnamefont
  {Aharonov}}, \bibinfo {author} {\bibfnamefont {David~Z.}\ \bibnamefont
  {Albert}}, \ and\ \bibinfo {author} {\bibfnamefont {Lev}\ \bibnamefont
  {Vaidman}},\ }\bibfield  {title} {\enquote {\bibinfo {title} {How the result
  of a measurement of a component of the spin of a spin-<i>1/2</i> particle can
  turn out to be 100},}\ }\href {\doibase 10.1103/PhysRevLett.60.1351}
  {\bibfield  {journal} {\bibinfo  {journal} {Phys. Rev. Lett.}\ }\textbf
  {\bibinfo {volume} {60}},\ \bibinfo {pages} {1351--1354} (\bibinfo {year}
  {1988})}\BibitemShut {NoStop}%
\bibitem [{\citenamefont {Lundeen}\ \emph {et~al.}(2011)\citenamefont
  {Lundeen}, \citenamefont {Sutherland}, \citenamefont {Patel}, \citenamefont
  {Stewrt},\ and\ \citenamefont {Bamber}}]{Lundeen}%
  \BibitemOpen
  \bibfield  {author} {\bibinfo {author} {\bibfnamefont {Jeff~S.}\ \bibnamefont
  {Lundeen}}, \bibinfo {author} {\bibfnamefont {Brandon}\ \bibnamefont
  {Sutherland}}, \bibinfo {author} {\bibfnamefont {Aabid}\ \bibnamefont
  {Patel}}, \bibinfo {author} {\bibfnamefont {Corey}\ \bibnamefont {Stewrt}}, \
  and\ \bibinfo {author} {\bibfnamefont {C.}~\bibnamefont {Bamber}},\
  }\bibfield  {title} {\enquote {\bibinfo {title} {Direct measurement of the
  quantum wavefunction},}\ }\href@noop {} {\bibfield  {journal} {\bibinfo
  {journal} {Nature}\ }\textbf {\bibinfo {volume} {474}},\ \bibinfo {pages}
  {188--191} (\bibinfo {year} {2011})}\BibitemShut {NoStop}%
\bibitem [{\citenamefont {Resch}\ \emph {et~al.}(2004)\citenamefont {Resch},
  \citenamefont {Lundeen},\ and\ \citenamefont {Steinberg}}]{Resch}%
  \BibitemOpen
  \bibfield  {author} {\bibinfo {author} {\bibfnamefont {J.~K.}\ \bibnamefont
  {Resch}}, \bibinfo {author} {\bibfnamefont {J.S.}\ \bibnamefont {Lundeen}}, \
  and\ \bibinfo {author} {\bibfnamefont {A.M.}\ \bibnamefont {Steinberg}},\
  }\bibfield  {title} {\enquote {\bibinfo {title} {Experimental realization of
  the quantum box problem},}\ }\href {\doibase
  http://dx.doi.org/10.1016/j.physleta.2004.02.042} {\bibfield  {journal}
  {\bibinfo  {journal} {Physics Letters A}\ }\textbf {\bibinfo {volume}
  {324}},\ \bibinfo {pages} {125 -- 131} (\bibinfo {year} {2004})}\BibitemShut
  {NoStop}%
\bibitem [{\citenamefont {Lundeen}\ and\ \citenamefont
  {Steinberg}(2009)}]{Hardy}%
  \BibitemOpen
  \bibfield  {author} {\bibinfo {author} {\bibfnamefont {J.~S.}\ \bibnamefont
  {Lundeen}}\ and\ \bibinfo {author} {\bibfnamefont {A.~M.}\ \bibnamefont
  {Steinberg}},\ }\bibfield  {title} {\enquote {\bibinfo {title} {Experimental
  joint weak measurement on a photon pair as a probe of hardy's paradox},}\
  }\href {\doibase 10.1103/PhysRevLett.102.020404} {\bibfield  {journal}
  {\bibinfo  {journal} {Phys. Rev. Lett.}\ }\textbf {\bibinfo {volume} {102}},\
  \bibinfo {pages} {020404} (\bibinfo {year} {2009})}\BibitemShut {NoStop}%
\bibitem [{\citenamefont {Lundeen}\ and\ \citenamefont
  {Bamber}(2012)}]{Lundeen2}%
  \BibitemOpen
  \bibfield  {author} {\bibinfo {author} {\bibfnamefont {Jeff~S.}\ \bibnamefont
  {Lundeen}}\ and\ \bibinfo {author} {\bibfnamefont {Charles}\ \bibnamefont
  {Bamber}},\ }\bibfield  {title} {\enquote {\bibinfo {title} {Procedure for
  direct measurement of general quantum states using weak measurement},}\
  }\href {\doibase 10.1103/PhysRevLett.108.070402} {\bibfield  {journal}
  {\bibinfo  {journal} {Phys. Rev. Lett.}\ }\textbf {\bibinfo {volume} {108}},\
  \bibinfo {pages} {070402} (\bibinfo {year} {2012})}\BibitemShut {NoStop}%
\bibitem [{\citenamefont {Howland}\ \emph {et~al.}(2014)\citenamefont
  {Howland}, \citenamefont {Lum},\ and\ \citenamefont {Howell}}]{Greg}%
  \BibitemOpen
  \bibfield  {author} {\bibinfo {author} {\bibfnamefont {Gregory~A.}\
  \bibnamefont {Howland}}, \bibinfo {author} {\bibfnamefont {Daniel~J.}\
  \bibnamefont {Lum}}, \ and\ \bibinfo {author} {\bibfnamefont {John~C.}\
  \bibnamefont {Howell}},\ }\bibfield  {title} {\enquote {\bibinfo {title}
  {Compressive wavefront sensing with weak values},}\ }\href {\doibase
  10.1364/OE.22.018870} {\bibfield  {journal} {\bibinfo  {journal} {Opt.
  Express}\ }\textbf {\bibinfo {volume} {22}},\ \bibinfo {pages} {18870--18880}
  (\bibinfo {year} {2014})}\BibitemShut {NoStop}%
\bibitem [{\citenamefont {Denkmayr}\ \emph {et~al.}(2014)\citenamefont
  {Denkmayr}, \citenamefont {Geppert}, \citenamefont {Sponar}, \citenamefont
  {Lemmel}, \citenamefont {Matzkin}, \citenamefont {Tollaksen},\ and\
  \citenamefont {Hasegawa}}]{QuantumCat}%
  \BibitemOpen
  \bibfield  {author} {\bibinfo {author} {\bibfnamefont {Tobias}\ \bibnamefont
  {Denkmayr}}, \bibinfo {author} {\bibfnamefont {Hermann}\ \bibnamefont
  {Geppert}}, \bibinfo {author} {\bibfnamefont {Stephan}\ \bibnamefont
  {Sponar}}, \bibinfo {author} {\bibfnamefont {Hartmut}\ \bibnamefont
  {Lemmel}}, \bibinfo {author} {\bibfnamefont {Alexandre}\ \bibnamefont
  {Matzkin}}, \bibinfo {author} {\bibfnamefont {Jeff}\ \bibnamefont
  {Tollaksen}}, \ and\ \bibinfo {author} {\bibfnamefont {Yuji}\ \bibnamefont
  {Hasegawa}},\ }\bibfield  {title} {\enquote {\bibinfo {title} {Observation of
  a quantum cheshire cat in a matter-wave interferometer experiment},}\
  }\href@noop {} {\bibfield  {journal} {\bibinfo  {journal} {Nature
  communications}\ }\textbf {\bibinfo {volume} {5}} (\bibinfo {year}
  {2014})}\BibitemShut {NoStop}%
\bibitem [{\citenamefont {Corrêa}\ \emph {et~al.}(2015)\citenamefont
  {Corrêa}, \citenamefont {Santos}, \citenamefont {Monken},\ and\
  \citenamefont {Saldanha}}]{CheshireCatBrazil}%
  \BibitemOpen
  \bibfield  {author} {\bibinfo {author} {\bibfnamefont {Raul}\ \bibnamefont
  {Corrêa}}, \bibinfo {author} {\bibfnamefont {Marcelo~França}\ \bibnamefont
  {Santos}}, \bibinfo {author} {\bibfnamefont {C~H}\ \bibnamefont {Monken}}, \
  and\ \bibinfo {author} {\bibfnamefont {Pablo~L}\ \bibnamefont {Saldanha}},\
  }\bibfield  {title} {\enquote {\bibinfo {title} {‘quantum cheshire cat’
  as simple quantum interference},}\ }\href
  {http://stacks.iop.org/1367-2630/17/i=5/a=053042} {\bibfield  {journal}
  {\bibinfo  {journal} {New Journal of Physics}\ }\textbf {\bibinfo {volume}
  {17}},\ \bibinfo {pages} {053042} (\bibinfo {year} {2015})}\BibitemShut
  {NoStop}%
\bibitem [{\citenamefont {Hosten}\ and\ \citenamefont {Kwiat}(2008)}]{Hosten}%
  \BibitemOpen
  \bibfield  {author} {\bibinfo {author} {\bibfnamefont {Onur}\ \bibnamefont
  {Hosten}}\ and\ \bibinfo {author} {\bibfnamefont {Paul}\ \bibnamefont
  {Kwiat}},\ }\bibfield  {title} {\enquote {\bibinfo {title} {Observation of
  the spin hall effect of light via weak measurements},}\ }\href {\doibase
  10.1126/science.1152697} {\bibfield  {journal} {\bibinfo  {journal}
  {Science}\ }\textbf {\bibinfo {volume} {319}},\ \bibinfo {pages} {787--790}
  (\bibinfo {year} {2008})},\ \Eprint
  {http://arxiv.org/abs/http://www.sciencemag.org/content/319/5864/787.full.pdf}
  {http://www.sciencemag.org/content/319/5864/787.full.pdf} \BibitemShut
  {NoStop}%
\bibitem [{\citenamefont {Dixon}\ \emph {et~al.}(2009)\citenamefont {Dixon},
  \citenamefont {Starling}, \citenamefont {Jordan},\ and\ \citenamefont
  {Howell}}]{Dixon2009}%
  \BibitemOpen
  \bibfield  {author} {\bibinfo {author} {\bibfnamefont {P.~Ben}\ \bibnamefont
  {Dixon}}, \bibinfo {author} {\bibfnamefont {David~J.}\ \bibnamefont
  {Starling}}, \bibinfo {author} {\bibfnamefont {Andrew~N.}\ \bibnamefont
  {Jordan}}, \ and\ \bibinfo {author} {\bibfnamefont {John~C.}\ \bibnamefont
  {Howell}},\ }\bibfield  {title} {\enquote {\bibinfo {title} {Ultrasensitive
  beam deflection measurement via interferometric weak value amplification},}\
  }\href {\doibase 10.1103/PhysRevLett.102.173601} {\bibfield  {journal}
  {\bibinfo  {journal} {Phys. Rev. Lett.}\ }\textbf {\bibinfo {volume} {102}},\
  \bibinfo {pages} {173601} (\bibinfo {year} {2009})}\BibitemShut {NoStop}%
\bibitem [{\citenamefont {Viza}\ \emph {et~al.}(2013)\citenamefont {Viza},
  \citenamefont {Mart\'{i}nez-Rinc\'{o}n}, \citenamefont {Howland},
  \citenamefont {Frostig}, \citenamefont {Shomroni}, \citenamefont {Dayan},\
  and\ \citenamefont {Howell}}]{Viza}%
  \BibitemOpen
  \bibfield  {author} {\bibinfo {author} {\bibfnamefont {Gerardo~I.}\
  \bibnamefont {Viza}}, \bibinfo {author} {\bibfnamefont {Juli\'{a}n}\
  \bibnamefont {Mart\'{i}nez-Rinc\'{o}n}}, \bibinfo {author} {\bibfnamefont
  {Gregory~A.}\ \bibnamefont {Howland}}, \bibinfo {author} {\bibfnamefont
  {Hadas}\ \bibnamefont {Frostig}}, \bibinfo {author} {\bibfnamefont {Itay}\
  \bibnamefont {Shomroni}}, \bibinfo {author} {\bibfnamefont {Barak}\
  \bibnamefont {Dayan}}, \ and\ \bibinfo {author} {\bibfnamefont {John~C.}\
  \bibnamefont {Howell}},\ }\bibfield  {title} {\enquote {\bibinfo {title}
  {Weak-values technique for velocity measurements},}\ }\href {\doibase
  10.1364/OL.38.002949} {\bibfield  {journal} {\bibinfo  {journal} {Opt.
  Lett.}\ }\textbf {\bibinfo {volume} {38}},\ \bibinfo {pages} {2949--2952}
  (\bibinfo {year} {2013})}\BibitemShut {NoStop}%
\bibitem [{\citenamefont {Egan}\ and\ \citenamefont {Stone}(2012)}]{Egan}%
  \BibitemOpen
  \bibfield  {author} {\bibinfo {author} {\bibfnamefont {Patrick}\ \bibnamefont
  {Egan}}\ and\ \bibinfo {author} {\bibfnamefont {Jack~A.}\ \bibnamefont
  {Stone}},\ }\bibfield  {title} {\enquote {\bibinfo {title} {Weak-value
  thermostat with 0.2\&\#xa0;mk precision},}\ }\href {\doibase
  10.1364/OL.37.004991} {\bibfield  {journal} {\bibinfo  {journal} {Opt.
  Lett.}\ }\textbf {\bibinfo {volume} {37}},\ \bibinfo {pages} {4991--4993}
  (\bibinfo {year} {2012})}\BibitemShut {NoStop}%
\bibitem [{\citenamefont {Starling}\ \emph
  {et~al.}(2010{\natexlab{a}})\citenamefont {Starling}, \citenamefont {Dixon},
  \citenamefont {Jordan},\ and\ \citenamefont {Howell}}]{StarlingFreq}%
  \BibitemOpen
  \bibfield  {author} {\bibinfo {author} {\bibfnamefont {David~J.}\
  \bibnamefont {Starling}}, \bibinfo {author} {\bibfnamefont {P.~Ben}\
  \bibnamefont {Dixon}}, \bibinfo {author} {\bibfnamefont {Andrew~N.}\
  \bibnamefont {Jordan}}, \ and\ \bibinfo {author} {\bibfnamefont {John~C.}\
  \bibnamefont {Howell}},\ }\bibfield  {title} {\enquote {\bibinfo {title}
  {Precision frequency measurements with interferometric weak values},}\ }\href
  {\doibase 10.1103/PhysRevA.82.063822} {\bibfield  {journal} {\bibinfo
  {journal} {Phys. Rev. A}\ }\textbf {\bibinfo {volume} {82}},\ \bibinfo
  {pages} {063822} (\bibinfo {year} {2010}{\natexlab{a}})}\BibitemShut
  {NoStop}%
\bibitem [{\citenamefont {Jayaswal}\ \emph {et~al.}(2014)\citenamefont
  {Jayaswal}, \citenamefont {Mistura},\ and\ \citenamefont
  {Merano}}]{jayaswal}%
  \BibitemOpen
  \bibfield  {author} {\bibinfo {author} {\bibfnamefont {G.}~\bibnamefont
  {Jayaswal}}, \bibinfo {author} {\bibfnamefont {G.}~\bibnamefont {Mistura}}, \
  and\ \bibinfo {author} {\bibfnamefont {M.}~\bibnamefont {Merano}},\
  }\bibfield  {title} {\enquote {\bibinfo {title} {Observing angular deviations
  in light-beam reflection via weak measurements},}\ }\href {\doibase
  10.1364/OL.39.006257} {\bibfield  {journal} {\bibinfo  {journal} {Opt.
  Lett.}\ }\textbf {\bibinfo {volume} {39}},\ \bibinfo {pages} {6257--6260}
  (\bibinfo {year} {2014})}\BibitemShut {NoStop}%
\bibitem [{\citenamefont {Li}\ \emph {et~al.}(2011)\citenamefont {Li},
  \citenamefont {Xu}, \citenamefont {Tang}, \citenamefont {Xu},\ and\
  \citenamefont {Guo}}]{Li}%
  \BibitemOpen
  \bibfield  {author} {\bibinfo {author} {\bibfnamefont {Chuan-Feng}\
  \bibnamefont {Li}}, \bibinfo {author} {\bibfnamefont {Xiao-Ye}\ \bibnamefont
  {Xu}}, \bibinfo {author} {\bibfnamefont {Jian-Shun}\ \bibnamefont {Tang}},
  \bibinfo {author} {\bibfnamefont {Jin-Shi}\ \bibnamefont {Xu}}, \ and\
  \bibinfo {author} {\bibfnamefont {Guang-Can}\ \bibnamefont {Guo}},\
  }\bibfield  {title} {\enquote {\bibinfo {title} {Ultrasensitive phase
  estimation with white light},}\ }\href {\doibase 10.1103/PhysRevA.83.044102}
  {\bibfield  {journal} {\bibinfo  {journal} {Phys. Rev. A}\ }\textbf {\bibinfo
  {volume} {83}},\ \bibinfo {pages} {044102} (\bibinfo {year}
  {2011})}\BibitemShut {NoStop}%
\bibitem [{\citenamefont {Starling}\ \emph
  {et~al.}(2010{\natexlab{b}})\citenamefont {Starling}, \citenamefont {Dixon},
  \citenamefont {Williams}, \citenamefont {Jordan},\ and\ \citenamefont
  {Howell}}]{StarlingPhase}%
  \BibitemOpen
  \bibfield  {author} {\bibinfo {author} {\bibfnamefont {David~J.}\
  \bibnamefont {Starling}}, \bibinfo {author} {\bibfnamefont {P.~Ben}\
  \bibnamefont {Dixon}}, \bibinfo {author} {\bibfnamefont {Nathan~S.}\
  \bibnamefont {Williams}}, \bibinfo {author} {\bibfnamefont {Andrew~N.}\
  \bibnamefont {Jordan}}, \ and\ \bibinfo {author} {\bibfnamefont {John~C.}\
  \bibnamefont {Howell}},\ }\bibfield  {title} {\enquote {\bibinfo {title}
  {Continuous phase amplification with a sagnac interferometer},}\ }\href
  {\doibase 10.1103/PhysRevA.82.011802} {\bibfield  {journal} {\bibinfo
  {journal} {Phys. Rev. A}\ }\textbf {\bibinfo {volume} {82}},\ \bibinfo
  {pages} {011802} (\bibinfo {year} {2010}{\natexlab{b}})}\BibitemShut
  {NoStop}%
\bibitem [{\citenamefont {Jordan}\ \emph {et~al.}(2014)\citenamefont {Jordan},
  \citenamefont {Mart\'{i}nez-Rinc\'{o}n},\ and\ \citenamefont
  {Howell}}]{Jordan2013}%
  \BibitemOpen
  \bibfield  {author} {\bibinfo {author} {\bibfnamefont {Andrew~N.}\
  \bibnamefont {Jordan}}, \bibinfo {author} {\bibfnamefont {Juli\'an}\
  \bibnamefont {Mart\'{i}nez-Rinc\'{o}n}}, \ and\ \bibinfo {author}
  {\bibfnamefont {John~C.}\ \bibnamefont {Howell}},\ }\bibfield  {title}
  {\enquote {\bibinfo {title} {Technical advantages for weak-value
  amplification: When less is more},}\ }\href {\doibase
  10.1103/PhysRevX.4.011031} {\bibfield  {journal} {\bibinfo  {journal} {Phys.
  Rev. X}\ }\textbf {\bibinfo {volume} {4}},\ \bibinfo {pages} {011031}
  (\bibinfo {year} {2014})}\BibitemShut {NoStop}%
\bibitem [{\citenamefont {Viza}\ \emph {et~al.}(2015)\citenamefont {Viza},
  \citenamefont {Mart\'{\i}nez-Rinc\'on}, \citenamefont {Alves}, \citenamefont
  {Jordan},\ and\ \citenamefont {Howell}}]{Viza2}%
  \BibitemOpen
  \bibfield  {author} {\bibinfo {author} {\bibfnamefont {Gerardo~I.}\
  \bibnamefont {Viza}}, \bibinfo {author} {\bibfnamefont {Juli\'an}\
  \bibnamefont {Mart\'{\i}nez-Rinc\'on}}, \bibinfo {author} {\bibfnamefont
  {Gabriel~B.}\ \bibnamefont {Alves}}, \bibinfo {author} {\bibfnamefont
  {Andrew~N.}\ \bibnamefont {Jordan}}, \ and\ \bibinfo {author} {\bibfnamefont
  {John~C.}\ \bibnamefont {Howell}},\ }\bibfield  {title} {\enquote {\bibinfo
  {title} {Experimentally quantifying the advantages of weak-value-based
  metrology},}\ }\href {\doibase 10.1103/PhysRevA.92.032127} {\bibfield
  {journal} {\bibinfo  {journal} {Phys. Rev. A}\ }\textbf {\bibinfo {volume}
  {92}},\ \bibinfo {pages} {032127} (\bibinfo {year} {2015})}\BibitemShut
  {NoStop}%
\bibitem [{\citenamefont {Nakamura}\ \emph {et~al.}(2012)\citenamefont
  {Nakamura}, \citenamefont {Nishizawa},\ and\ \citenamefont
  {Fujimoto}}]{Nakamura}%
  \BibitemOpen
  \bibfield  {author} {\bibinfo {author} {\bibfnamefont {Kouji}\ \bibnamefont
  {Nakamura}}, \bibinfo {author} {\bibfnamefont {Atsushi}\ \bibnamefont
  {Nishizawa}}, \ and\ \bibinfo {author} {\bibfnamefont {Masa-Katsu}\
  \bibnamefont {Fujimoto}},\ }\bibfield  {title} {\enquote {\bibinfo {title}
  {Evaluation of weak measurements to all orders},}\ }\href {\doibase
  10.1103/PhysRevA.85.012113} {\bibfield  {journal} {\bibinfo  {journal} {Phys.
  Rev. A}\ }\textbf {\bibinfo {volume} {85}},\ \bibinfo {pages} {012113}
  (\bibinfo {year} {2012})}\BibitemShut {NoStop}%
\bibitem [{\citenamefont {Pang}\ \emph {et~al.}(2012)\citenamefont {Pang},
  \citenamefont {Wu},\ and\ \citenamefont {Chen}}]{PanOrthogonal}%
  \BibitemOpen
  \bibfield  {author} {\bibinfo {author} {\bibfnamefont {Shengshi}\
  \bibnamefont {Pang}}, \bibinfo {author} {\bibfnamefont {Shengjun}\
  \bibnamefont {Wu}}, \ and\ \bibinfo {author} {\bibfnamefont {Zeng-Bing}\
  \bibnamefont {Chen}},\ }\bibfield  {title} {\enquote {\bibinfo {title} {Weak
  measurement with orthogonal preselection and postselection},}\ }\href
  {\doibase 10.1103/PhysRevA.86.022112} {\bibfield  {journal} {\bibinfo
  {journal} {Phys. Rev. A}\ }\textbf {\bibinfo {volume} {86}},\ \bibinfo
  {pages} {022112} (\bibinfo {year} {2012})}\BibitemShut {NoStop}%
\bibitem [{\citenamefont {Koike}\ and\ \citenamefont {Tanaka}(2011)}]{Koike}%
  \BibitemOpen
  \bibfield  {author} {\bibinfo {author} {\bibfnamefont {Tatsuhiko}\
  \bibnamefont {Koike}}\ and\ \bibinfo {author} {\bibfnamefont {Saki}\
  \bibnamefont {Tanaka}},\ }\bibfield  {title} {\enquote {\bibinfo {title}
  {Limits on amplification by aharonov-albert-vaidman weak measurement},}\
  }\href {\doibase 10.1103/PhysRevA.84.062106} {\bibfield  {journal} {\bibinfo
  {journal} {Phys. Rev. A}\ }\textbf {\bibinfo {volume} {84}},\ \bibinfo
  {pages} {062106} (\bibinfo {year} {2011})}\BibitemShut {NoStop}%
\bibitem [{\citenamefont {Lorenzo}(2014)}]{DiLorenzo}%
  \BibitemOpen
  \bibfield  {author} {\bibinfo {author} {\bibfnamefont {Antonio~Di}\
  \bibnamefont {Lorenzo}},\ }\bibfield  {title} {\enquote {\bibinfo {title}
  {Weak values and weak coupling maximizing the output of weak measurements},}\
  }\href {\doibase http://dx.doi.org/10.1016/j.aop.2014.03.007} {\bibfield
  {journal} {\bibinfo  {journal} {Annals of Physics}\ }\textbf {\bibinfo
  {volume} {345}},\ \bibinfo {pages} {178 -- 189} (\bibinfo {year}
  {2014})}\BibitemShut {NoStop}%
\bibitem [{\citenamefont {Hariharan}\ and\ \citenamefont
  {Roy}(1992)}]{Hariharan}%
  \BibitemOpen
  \bibfield  {author} {\bibinfo {author} {\bibfnamefont {P.}~\bibnamefont
  {Hariharan}}\ and\ \bibinfo {author} {\bibfnamefont {Maitreyee}\ \bibnamefont
  {Roy}},\ }\bibfield  {title} {\enquote {\bibinfo {title} {A geometric-phase
  interferometer},}\ }\href {\doibase 10.1080/09500349214551881} {\bibfield
  {journal} {\bibinfo  {journal} {Journal of Modern Optics}\ }\textbf {\bibinfo
  {volume} {39}},\ \bibinfo {pages} {1811--1815} (\bibinfo {year}
  {1992})}\BibitemShut {NoStop}%
\bibitem [{\citenamefont {Loredo}\ \emph {et~al.}(2009)\citenamefont {Loredo},
  \citenamefont {Ort\'{\i}z}, \citenamefont {Weing\"artner},\ and\
  \citenamefont {De~Zela}}]{Loredo}%
  \BibitemOpen
  \bibfield  {author} {\bibinfo {author} {\bibfnamefont {J.~C.}\ \bibnamefont
  {Loredo}}, \bibinfo {author} {\bibfnamefont {O.}~\bibnamefont {Ort\'{\i}z}},
  \bibinfo {author} {\bibfnamefont {R.}~\bibnamefont {Weing\"artner}}, \ and\
  \bibinfo {author} {\bibfnamefont {F.}~\bibnamefont {De~Zela}},\ }\bibfield
  {title} {\enquote {\bibinfo {title} {Measurement of pancharatnam's phase by
  robust interferometric and polarimetric methods},}\ }\href {\doibase
  10.1103/PhysRevA.80.012113} {\bibfield  {journal} {\bibinfo  {journal} {Phys.
  Rev. A}\ }\textbf {\bibinfo {volume} {80}},\ \bibinfo {pages} {012113}
  (\bibinfo {year} {2009})}\BibitemShut {NoStop}%
\bibitem [{\citenamefont {Howell}\ \emph {et~al.}(2010)\citenamefont {Howell},
  \citenamefont {Starling}, \citenamefont {Dixon}, \citenamefont {Vudyasetu},\
  and\ \citenamefont {Jordan}}]{JohnQM}%
  \BibitemOpen
  \bibfield  {author} {\bibinfo {author} {\bibfnamefont {John~C.}\ \bibnamefont
  {Howell}}, \bibinfo {author} {\bibfnamefont {David~J.}\ \bibnamefont
  {Starling}}, \bibinfo {author} {\bibfnamefont {P.~Ben}\ \bibnamefont
  {Dixon}}, \bibinfo {author} {\bibfnamefont {Praveen~K.}\ \bibnamefont
  {Vudyasetu}}, \ and\ \bibinfo {author} {\bibfnamefont {Andrew~N.}\
  \bibnamefont {Jordan}},\ }\bibfield  {title} {\enquote {\bibinfo {title}
  {Interferometric weak value deflections: Quantum and classical treatments},}\
  }\href {\doibase 10.1103/PhysRevA.81.033813} {\bibfield  {journal} {\bibinfo
  {journal} {Phys. Rev. A}\ }\textbf {\bibinfo {volume} {81}},\ \bibinfo
  {pages} {033813} (\bibinfo {year} {2010})}\BibitemShut {NoStop}%
\bibitem [{\citenamefont {Wu}\ and\ \citenamefont
  {Li}(2011)}]{ShengOrthogonal}%
  \BibitemOpen
  \bibfield  {author} {\bibinfo {author} {\bibfnamefont {Shengjun}\
  \bibnamefont {Wu}}\ and\ \bibinfo {author} {\bibfnamefont {Yang}\
  \bibnamefont {Li}},\ }\bibfield  {title} {\enquote {\bibinfo {title} {Weak
  measurements beyond the aharonov-albert-vaidman formalism},}\ }\href
  {\doibase 10.1103/PhysRevA.83.052106} {\bibfield  {journal} {\bibinfo
  {journal} {Phys. Rev. A}\ }\textbf {\bibinfo {volume} {83}},\ \bibinfo
  {pages} {052106} (\bibinfo {year} {2011})}\BibitemShut {NoStop}%
\bibitem [{Note1()}]{Note1}%
  \BibitemOpen
  \bibinfo {note} {We remind the reader that the data presented for the
  concatenated case is one of the two optimized cases (rows 3 and 4) where
  there is an enhancement in the signal, i.e. row 3 in Table~\ref
  {tab:table1}.}\BibitemShut {Stop}%
\bibitem [{\citenamefont {Pang}\ and\ \citenamefont {Brun}(2014)}]{Brun}%
  \BibitemOpen
  \bibfield  {author} {\bibinfo {author} {\bibfnamefont {S}~\bibnamefont
  {Pang}}\ and\ \bibinfo {author} {\bibfnamefont {T~A}\ \bibnamefont {Brun}},\
  }\bibfield  {title} {\enquote {\bibinfo {title} {Improving the precision of
  weak measurements by postselection},}\ }\href@noop {} {\bibfield  {journal}
  {\bibinfo  {journal} {arXiv preprint 1409.2567}\ } (\bibinfo {year}
  {2014})}\BibitemShut {NoStop}%
\bibitem [{\citenamefont {Mart\'{\i}nez-Rinc\'on}\ \emph
  {et~al.}(2016)\citenamefont {Mart\'{\i}nez-Rinc\'on}, \citenamefont {Liu},
  \citenamefont {Viza},\ and\ \citenamefont {Howell}}]{Julian}%
  \BibitemOpen
  \bibfield  {author} {\bibinfo {author} {\bibfnamefont {Juli\'an}\
  \bibnamefont {Mart\'{\i}nez-Rinc\'on}}, \bibinfo {author} {\bibfnamefont
  {Wei-Tao}\ \bibnamefont {Liu}}, \bibinfo {author} {\bibfnamefont
  {Gerardo~I.}\ \bibnamefont {Viza}}, \ and\ \bibinfo {author} {\bibfnamefont
  {John~C.}\ \bibnamefont {Howell}},\ }\bibfield  {title} {\enquote {\bibinfo
  {title} {Can anomalous amplification be attained without postselection?}}\
  }\href {\doibase 10.1103/PhysRevLett.116.100803} {\bibfield  {journal}
  {\bibinfo  {journal} {Phys. Rev. Lett.}\ }\textbf {\bibinfo {volume} {116}},\
  \bibinfo {pages} {100803} (\bibinfo {year} {2016})}\BibitemShut {NoStop}%
\bibitem [{\citenamefont {Liu}\ \emph {et~al.}()\citenamefont {Liu},
  \citenamefont {Mart\'{\i}nez-Rinc\'on}, \citenamefont {Viza},\ and\
  \citenamefont {Howell}}]{Weitao}%
  \BibitemOpen
  \bibfield  {author} {\bibinfo {author} {\bibfnamefont {Wei-Tao}\ \bibnamefont
  {Liu}}, \bibinfo {author} {\bibfnamefont {Juli\'an}\ \bibnamefont
  {Mart\'{\i}nez-Rinc\'on}}, \bibinfo {author} {\bibfnamefont
  {Gerardo~Iv{\'a}n}\ \bibnamefont {Viza}}, \ and\ \bibinfo {author}
  {\bibfnamefont {John~C.}\ \bibnamefont {Howell}},\ }\href@noop {} {\bibinfo
  {journal} {in preparation}\ }\BibitemShut {NoStop}%
\bibitem [{\citenamefont {Mart\'{\i}nez-Rinc\'on}\ \emph
  {et~al.}()\citenamefont {Mart\'{\i}nez-Rinc\'on}, \citenamefont {Chen},\ and\
  \citenamefont {Howell}}]{Zekai}%
  \BibitemOpen
\bibfield  {journal} {  }\bibfield  {author} {\bibinfo {author} {\bibfnamefont
  {Juli\'an}\ \bibnamefont {Mart\'{\i}nez-Rinc\'on}}, \bibinfo {author}
  {\bibfnamefont {Zekai}\ \bibnamefont {Chen}}, \ and\ \bibinfo {author}
  {\bibfnamefont {John~C.}\ \bibnamefont {Howell}},\ }\href@noop {} {\bibinfo
  {journal} {in preparation}\ }\BibitemShut {NoStop}%
\end{thebibliography}

%

\appendix

\onecolumngrid

\section{Quarter-Half-Quarter: Pancharatnam-Berry phase}\label{App:QHQ}

Inside the Sagnac interferometer the polarization dependent phase is controlled by the wave plate combination of quarter, half, and quarter wave plates. The quarter wave plates are set to $\pm45^\circ$ denoted as the $\hat{Q}$ matrices and the half wave plate is denoted by the $\hat{H}$ matrix. We denote the product of the three wave plate combination as the $\hat{C}$ matrix. We will represent the wave plate matrices in the Jones matrix formalism in the polarization basis of $H$ and $V$.

\begin{equation}
\begin{split}
\hat{C}_{\ket{\circlearrowright},\ket{\circlearrowleft}}\left(\pm\frac{\phi}{4}\right) &= \hat{Q}\left(45^\circ\right)\hat{H}\left(\pm\frac{\phi}{4}\right)\hat{Q}\left(-45^\circ\right)\\
&=\frac{1}{2}
\begin{pmatrix}
1 & -i\\
-i & 1
\end{pmatrix}
\begin{pmatrix}
\cos(\phi/2) & \pm\sin(\phi/2)\\
\pm\sin(\phi/2) & -\cos(\phi/2)
\end{pmatrix}
\begin{pmatrix}
1 & i\\
i & 1
\end{pmatrix}\\
&=
\begin{pmatrix}
0 & -e^{\mp i\phi/2}\\
e^{\pm i\phi/2} & 0
\end{pmatrix}
\end{split}
\end{equation}
We note that this configuration of wave plates gives a geometric phase that depends on the polarization of the beam.

We note the symmetry in $\hat{C}(\pm\phi/4)$ is broken by the beam propagation direction either clockwise ($\ket{\circlearrowright}$) with $+\phi/4$ or counter clockwise ($\ket{\circlearrowleft}$) with $-\phi/4$ as the half-wave plate angle. The wave plate combination $\hat{C}(\pm\phi/4)$ changes the state as follows: $\hat{C}(\phi/4)\ket{\circlearrowright}
\otimes\ket{H}=e^{i\phi/2}\ket{\circlearrowright}\otimes\ket{V}$, $\hat{C}(-\phi/4)\ket{\circlearrowleft}\otimes\ket{H}
=e^{-i\phi/2}\ket{\circlearrowleft}\otimes\ket{V}$, $\hat{C}(\phi/4)\ket{\circlearrowright} \otimes\ket{V}=-e^{-i\phi/2}\ket{\circlearrowright}
\otimes\ket{H}$ and $\hat{C}(-\phi/4)\ket{\circlearrowleft}\otimes\ket{V}
=-e^{i\phi/2}\ket{\circlearrowleft}\otimes\ket{H}$.

\section{Weak-value Quantum Description}\label{App:WVQM}
The preparation of state for this experiment deals with joint space between the which-path and polarization degree of freedom. The input state is first linearly polarized to the anti-diagonal state $\frac{1}{\sqrt{2}}(\ket{H}-\ket{V})$. Then the state enters the beam splitter of the Sagnac interferometer. We define the state after the beam splitter as
\begin{equation}
\ket{\xi}=\frac{1}{2}(\ket{\circlearrowright}+i\ket{\circlearrowleft})\otimes(\ket{H}-\ket{V})).
\end{equation}
To write the input state before the interaction we include the polarization dependent phase $\phi/2$ described in Appendix~\ref{App:QHQ}. 

\begin{equation}
\begin{split}
\ket{\varphi}_1 &=\hat{C}\left(\pm\phi/4\right)\ket{\xi}\\
&=\frac{1}{2}\left((\ket{\circlearrowright} e^{i\phi/2}+i\ket{\circlearrowleft}
e^{-i\phi/2})\otimes\ket{V}+(\ket{\circlearrowright} e^{-i\phi/2}+i\ket{\circlearrowleft}e^{i\phi/2})
\otimes\ket{H}\right).
\end{split}
\end{equation}
The interferometer has an interaction given by $\hat{U}=\exp(ik\hat{A}\hat{x})$ such that the transverse momentum $k$ is coupled by the ancillary operator $\hat{A}$ to the meter $\hat{x}$. The ancillary system operator $\hat{A}$ is given in the which-path basis by $\hat{A} = \ket{\circlearrowright}\bra{\circlearrowright}-\ket{\circlearrowleft}\bra{\circlearrowleft}$.  The parameter of interest is the transverse momentum kick $k$ that the beam of light receives on the reflected port of the beam splitter.
Then the first postselection is conducted with spatial interference. The postselection is nearly orthogonal to the input state and is given by
\begin{equation}
\ket{\varphi}_2=\frac{1}{\sqrt{2}}(\ket{\circlearrowright}-i\ket{\circlearrowleft}).
\end{equation}

We assume the interaction is weak such that we can expand to $\mathcal{O}(k^1)$ and can define a weak-value for both horizontal and vertical polarizations. The postselection is only for the spatial degree of freedom so we have two weak-values given by
\begin{subequations}
\begin{equation}
A^H_{wv}=\frac{\bra{H}\braket{\varphi_2|\hat{A}|\varphi_1}}{\bra{H}\braket{\varphi_2|\varphi_1}}=i\cot(\phi/2),
\end{equation}
and
\begin{equation}
A^V_{wv}=\frac{\bra{V}\braket{\varphi_2|\hat{A}|\varphi_1}}{\bra{V}\braket{\varphi_2|\varphi_1}}=-i\cot(\phi/2).
\end{equation}
\end{subequations}
We have two weak-values of opposite signs, thus the separation between the two polarization components becomes $2|\delta x_s|=4|k\sigma^2\cot(\phi/2)|$ and the signal on the detector is null. Then we introduce a second postselection in the polarization basis.

The second postselection is through polarization interference, and the postselected state is given by
\begin{equation}
\ket{\varphi}_3=\frac{1}{\sqrt{2}}\left[(\sin\theta-\cos\theta)\ket{H}-(\sin\theta+\cos\theta)
\ket{V}\right].
\end{equation}
The angle $\theta$ is a small angle that determines the orthogonality between the pre- and postselections in the polarization basis.
With the concatenated postselection we have a total effective weak-value given by
\begin{equation}
A^C_{wv}=\frac{\bra{\varphi_3}\braket{\varphi_2|\hat{A}|\varphi_1}}{\bra{\varphi_3}\braket{\varphi_2|\varphi_1}}=i\cot(\phi/2)\cot(\theta).
\end{equation}
With this concatenated configuration we amplify the visibility of the weak-value and improve the contrast of the spatial interference by adding the polarization degree of freedom.


\end{document}